\documentclass[twocolumn,twoside,slac_two]{revtex4}
\usepackage{graphicx}
\usepackage{fancyhdr}
\begin{document}

\title{Prompt optical observations of Fermi-LAT bursts and GRB 090902B}

\author{S. B. Pandey, Carl W. Akerlof, W. Zheng and F. Yuan}
\affiliation{Randall Laboratory of Physics, University of Michigan, Ann Arbor, Michigan, 48109-1040, USA}

\begin{abstract}

Observations of high energy emission from gamma-ray bursts (GRBs) constrain the extreme physical 
conditions associated with these energetic cosmic explosions. The Large Area Telescope (LAT) 
onboard the Fermi Gamma-ray Space Telescope, a pair conversion telescope, observes energetic quanta 
from 30 MeV to $>$ 300 GeV. Synergy of the LAT with the Gamma-ray Burst Monitor (GBM) enlarges the 
energy coverage to $\sim$7.5 decades, very useful for studying the GRB emission itself. Prompt 
optical observations and their possible correlations with photon emission at LAT energies help our 
understanding of the physical mechanisms behind these events. The prompt response times and large 
fields of of the ROTSE-III telescopes make afterglow observations possible for Fermi bursts with 
$\sim$1 degree localized errors. As an example, GRB 090902B, was observed starting $\sim$4803 s 
after the burst. This is the earliest ground-based optical detection ever made for long-duration 
bursts sensed by the LAT. The ROTSE detection classifies the optical afterglow of GRB 090902B as 
one of the brightest.

\end{abstract}

\maketitle

\thispagestyle{fancy}

\section{LAT GRBs and their importance:}
The Fermi Gamma-ray Space Telescope has opened a new high-energy window in the study of GRBs. The 
understanding about the origin of high energy emission from GRBs is quite limited due to both the 
small number of bursts with known high energy photons and the small number of quanta that were 
detected in such cases, specially by the EGRET on board CGRO~\cite{Dingus1995}. Observations of 
GRBs at the LAT energies will be able to not only constrain the nature of the prompt emission but 
will also help towards understanding the underlying radiation mechanism and the outflow 
composition~\cite{Band2009}. There are a dozen GRBs which have been seen at LAT energies 
though more than $\sim$ 300 GRBs have been detected by the GBM so far. For 6 of these bursts, the LAT 
boresight angle is $<$ 65 degrees. Optical afterglows have been observed for 7 of these cases 
starting from $\sim$ 16 hours after the burst and redshift values determined. So far, no prompt 
optical observations have been obtained for any one of the LAT-detected GRBs. In the past, near 
contemporaneous optical and gamma-ray observations of couple of GRBs have helped constrain 
the nature of prompt emission of these energetic cosmic explosions~\cite{Akerlof1999, Vestrand2006}.   

\section{GRB localization errors and their brightness:}

The accuracy of localization of GRBs is not only a function of the detection technique used but also 
the brightness of the source at that energy. The onboard GBM has an angular resolution of 10 
degrees whereas the LAT angular resolution is considerably better. The GRB localization errors 
computed by the method described in~\cite{Akerlof2006} are shown below in the left and right panels 
of Figure 1 for the expected number of photons and the GRB fluences with energies $>$ 30 MeV 
respectively. For comparison, vertical lines show the bounds for some typical instruments that have 
been used for rapid observations of GRB afterglows. A number of effects have been ignored such 
as the finite energy resolution of the calorimeter, degradation of the PSF for off-axis photons and 
the influence of PSF non-Gaussian tails. All of these will make the localization error larger. For 
this reason, the dotted lines show the effects of a reasonably optimistic degradation factor of 1.5.

Based on the simulated bursts, the localization accuracy was computed for both onboard and on-ground 
as a function of the expected number of counts at LAT frequencies~\cite{Band2009} and are shown below 
respectively in the left and right panels of Figure 2. The results presented above in Figures 1 are 
similar to those presented in Figure 2. From Figure 2, it is clear that the number of identified 
GRBs and their localization accuracies are improved by the ground-based analysis but at the cost of 
considerable delay. These results further indicate the importance of ground-based robotic optical 
telescopes that can obtain early optical observations.

\begin{figure*}
\centering
\includegraphics[width=75mm]{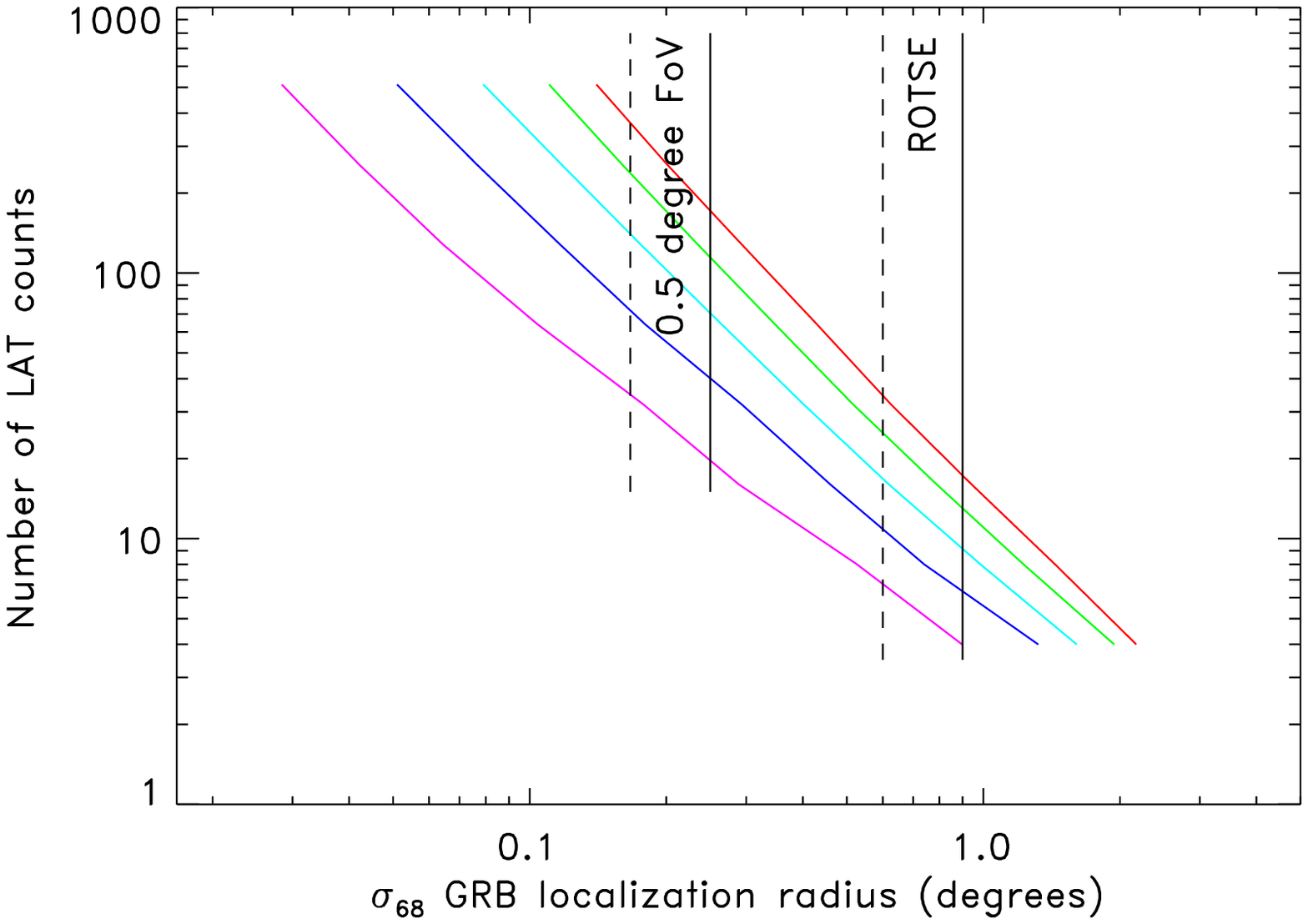}
\includegraphics[width=75mm]{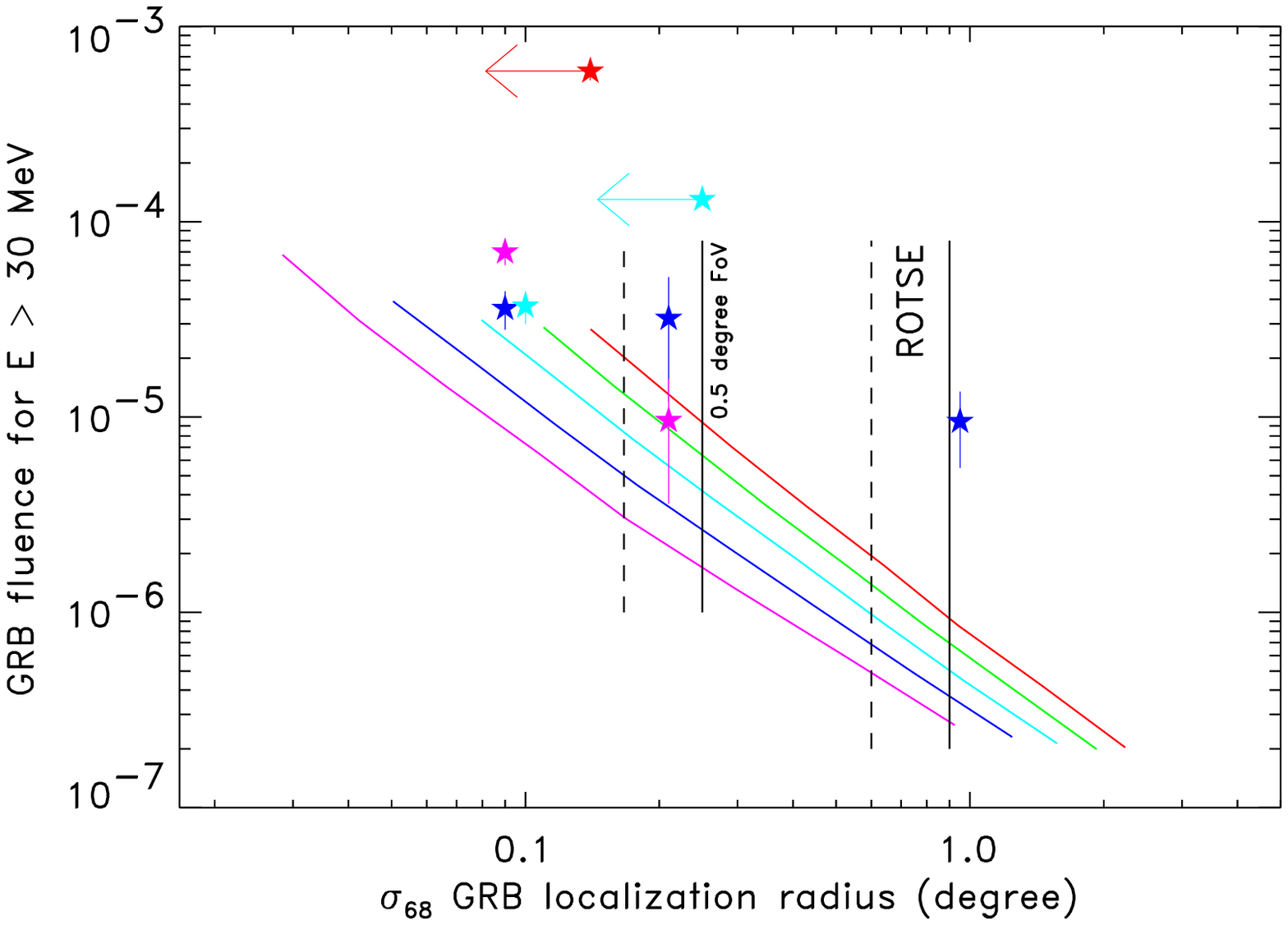}
\caption{The number of required LAT-detected photons as a function of the median GRB localization 
error is shown in the left panel. The five lines indicate the behaviour for the spectarl index 
$\beta$ = -2.00, -2.25, -2.50, -2.75 and -3.00, colour-coded from the top to bottom with an 
increasing vlaues of $\beta$. The vertical lines show the cutoffs for instruments with 0.5 degree 
diameter FoV and ROTSE-III. In the right panel, the required GRB fluence for E $>$ 30 MeV as a 
function of the median GRB localization error with five lines as described for the left 
panel of the Figure. Over plotted stars (colour coded to the nearest value of $\beta$ determined for 
GRBs listed in Table 1) represent the 8 GRBs tabulated in Table 1. It is clear that most of 
such bursts would have been suitable for ground-bases telescopes like ROTSE with the given LAT 
triggers.
}\label{Figure1}
\end{figure*}

\begin{figure*}
\centering
\includegraphics[width=75mm]{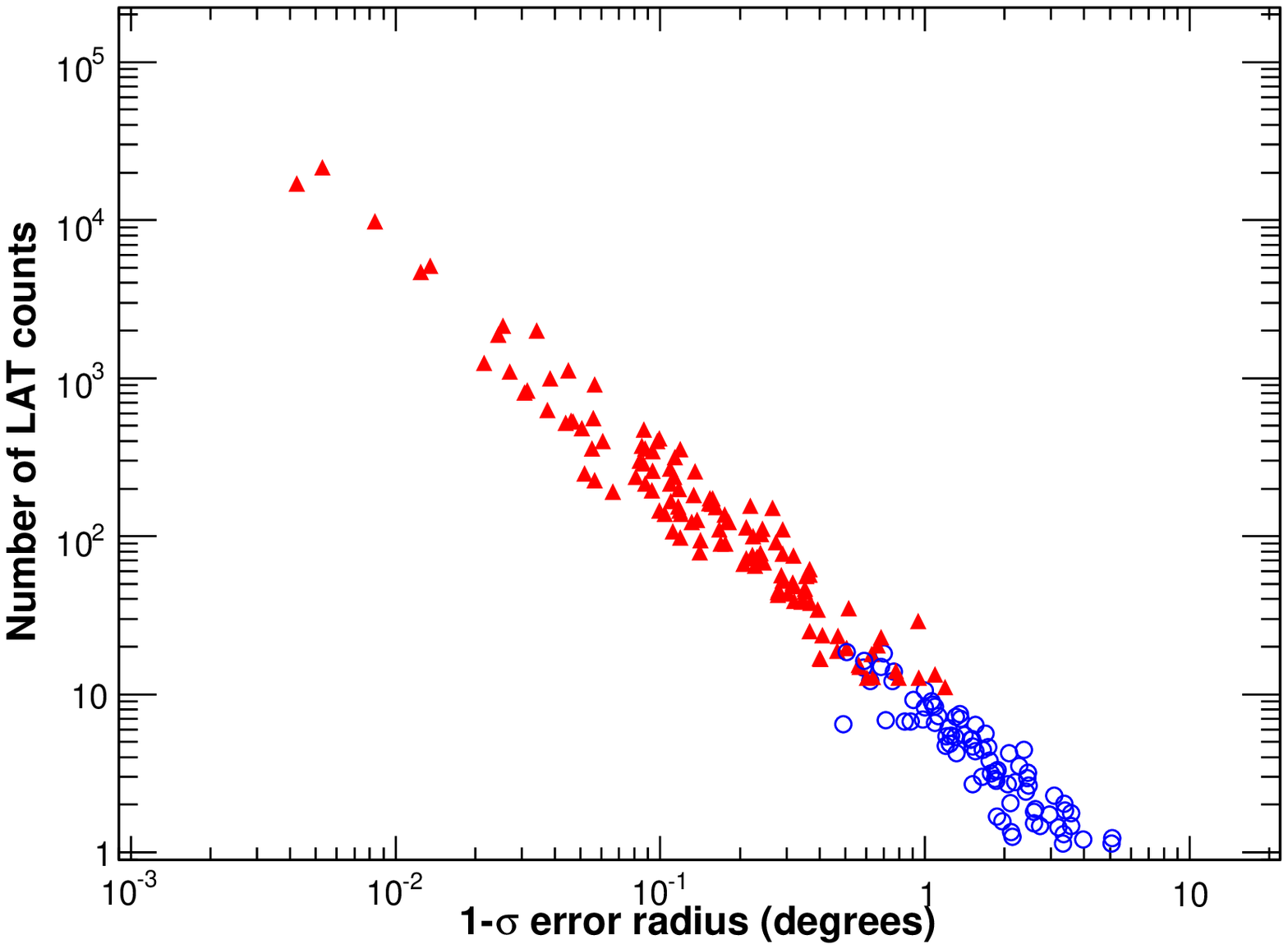}
\includegraphics[width=75mm]{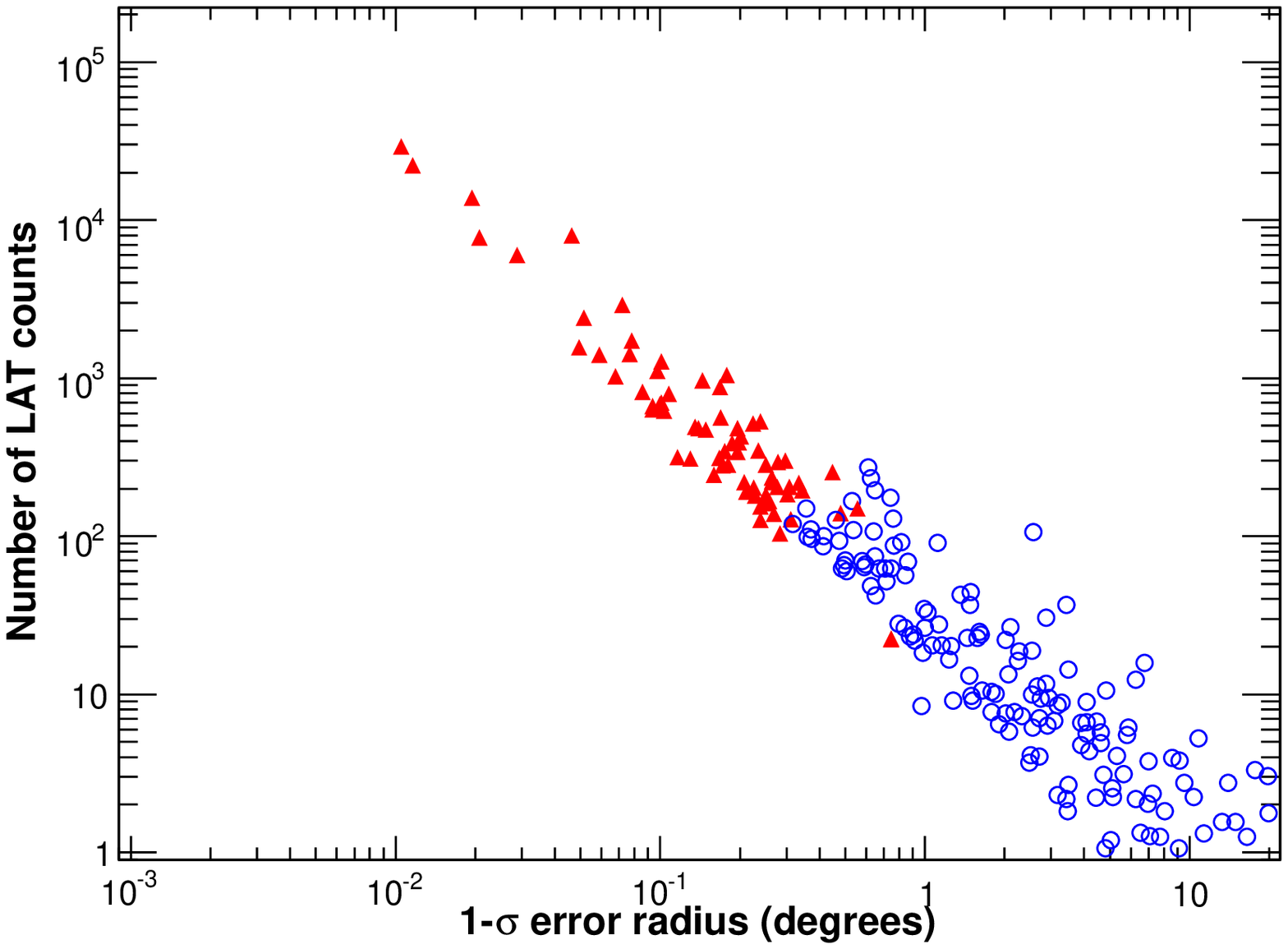}
\caption{The localization accuracy of LAT onboard and on-ground along with expected number of 
the LAT photons are plotted in the left and right panels (The figures borrowed from~\cite{Band2009}). In both the plots, the red triangles denote detected bursts and the open blue circles show undetected bursts.
}\label{Figure2}
\end{figure*}

\section{Optical Facilities and the LAT GRBs:}

Extensive multi-wavelength observations of the LAT-detected GRBs (except bursts expected to be 
detected both by Swift/BAT and Fermi/LAT, for example GRB 090510) will only be possible if 
ground-based instruments identify these events efficiently. A dozen or so events per year with 100 
LAT photons or more could be found by instruments with fields-of-view of 15 arcmin diameters for 
$\beta$ = -2.00 and 24 arcmin diameters for $\beta$ = -2.25. As shown in Figure 3 below, this 
requirement eliminates 61\% (for $\beta$ = -2.00) to 76\% (for $\beta$ = -2.25) of the observatories 
that have been reporting observations via GCN recently. Without the wider FoV systems such as RAPTOR, 
Super-LOTIS and ROTSE-III, the total discovery rate of such GRB afterglows may be restricted to 
3-4 events per year. If Nature is uncooperative, this could be considerably less. With a FoV of 
110'$\times$110' and global coverage, ROTSE-III is much better matched to a broader range of 
spectral indices and fluences.

\section{ROTSE Observations of the Fermi bursts:}

The ROTSE-III telescopes have demonstrated their suitability for the near-simultaneous GRB 
observations at optical frequencies. Since launch of the Fermi, ROTSE-III 
has responded to many of the bursts commonly seen by Swift/BAT and Fermi/GBM and the 
near-simultaneous observations were made for 3 bursts, namely GRB 080810, GRB 080928 
(~\cite{Rossi2009}, the poster presented in this meeting) and GRB 090618.  There is no information 
about the LAT data for these 3 bursts. The LAT boresight angle is 61 degrees for 
GRB 080810 and is 133 degrees for GRB 090618. GRB 090902B is the first LAT burst observed by 
ROTSE telescopes. The detail of observations and the properties of the burst are described below.

\subsection{GRB 090902B:}

The receipt of a ground-corrected GBM trigger~\cite{Bissaldi2009} with a 1 degree nominal location 
error initiated an observing sequence for ROTSE-IIIa, located at the Siding Spring Observatory in 
Australia. The telescope began taking three sets of thirty 20-s images, around the GBM estimated 
location. Only the third set, starting $\sim$ 80 minutes after the burst, with R.A. = 17h 38m 13s 
and Dec. = +27d 30' 59" covered the XRT burst location later identified as R.A. = 17h 39m 45.26s 
and Dec. = +27d 19' 28.1"~\cite{Kennea2009}. A substantial fraction of the delay was imposed by the 
interpolation of a previous observation request for an unrelated field. The images were processed 
using the standard ROTSE-III software pipeline and the unfiltered magnitudes were calibrated to 
$R-$band with respect to nearby USNO B1.0 stars using the method described in ~\cite{Quimby2006} 
and reported in ~\cite{Pandey2009}.

Based on the statistical analysis of the total ensemble of 30 ROTSE-III measurements, we estimate 
a spurious detection probability of less than 1\%. Combined with the spatial localization constraint, the probability of a false identification is less than 1$\times10^{-4}$. The comparison of the 
optical brightness of GRB 090902B (extrapolating to 1000 sec after the burst) with the apparent 
optical brightness probability distribution function of well-observed afterglows~\cite{Akerlof2007} 
places this event among the top 3\% of the brightest bursts observed so far. The burst is among the 
top 5\% of bright bursts in the comparison of the brightness at 1000 sec with the compilation of 
light-curves of optical afterglows of pre-Swift and Swift-era GRBs with known 
redshifts~\cite{Kann2007}.

\begin{figure}[t]
\centering
\includegraphics[width=75mm]{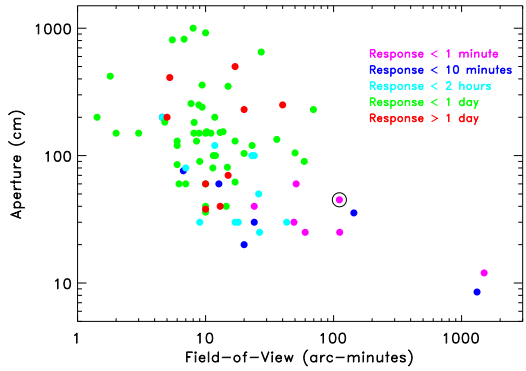}
\caption{Scatter plot of 82 optical facilities producing GCN messages during 2005-2008. The colour 
code designates the minimum response time for the particular system. The parameters for the 
ROTSE-III instruments are identified by the circumscribed circle. The aperture (in cm) and 
field-of-view (in arcmin) were obtained from online documentation or relevant publications.
} \label{Figure3}
\end{figure}

\begin{figure}
\centering
\includegraphics[width=75mm]{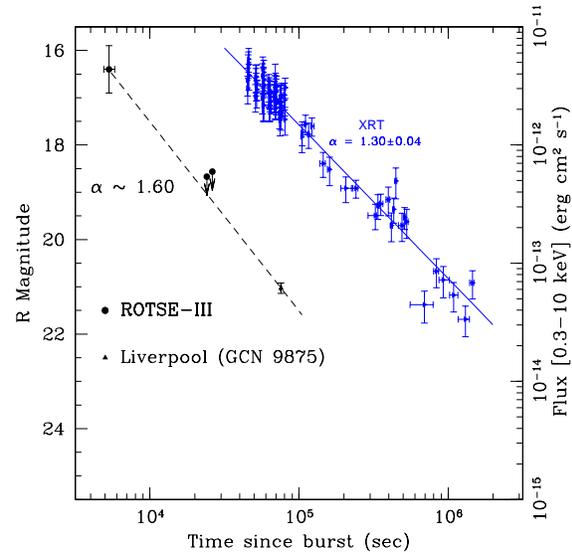}
\caption{The afterglow lightcurve of GRB 090902B at optical and XRT frequencies. 
The optical observations seems decaying similar to LAT one~\cite{Abdo2009} whereas XRT lightcurve 
decay rate is shallower. ROTSE observations are the earliest one ever taken for the long-duration 
GRBs by LAT.
}\label{Figure4}
\end{figure}

\subsection{Optical brightness of other LAT bursts:}

Optical afterglows of other LAT-detected long-duration GRBs, except GRB 090902B, were obtained 
starting $\sim$ 16 to 26 hours post-burst\footnote{\emph{http://lyra.berkeley.edu/grbox/grbox.php}} and 
the redshift values were determined for most of 
them\footnote{\emph{http://www.mpe.mpg.de/~jcg/grbgen.html}}. Comparison of the optical 
brightness of these LAT-detected bursts with that of other well observed GRBs in the time-scales of 
16 to 26 hours post-burst, indicate that they have a typical brightness as seen in the cases of 
pre-Swift and Swift GRBs~\cite{Kann2007}. The optical afterglow of GRB 090926 was exceptionally 
bright even at these later epochs~\cite{Haislip2009}.

\begin{table*}[t]
\begin{center}
\scriptsize
\caption{List of the 8 LAT-detected GRBs (Col. 1) with boresight angle $<$ 65 degrees (Col. 2) 
and/or known redshifts (Col. 3). The Spectral index (Col. 4), the LAT fluence 
(Col. 5)~\cite{Ghiselini2009} and the 1-sigme localization errors (Col 6.) of these GRBs are over 
plotted in Figure 1.
For more details, http://fermi.gsfc.nasa.gov/ssc/resources/observations/grbs/grb$\_$table}
\vspace{0.6cm}
\begin{tabular}{|l|c|c|c|c|c|}
\hline \textbf{Name of GRB}&\textbf{Boresight angle}&\textbf{Redshift}&\textbf{Spectral index}&\textbf{LAT fluence}&\textbf{1-sigma LAT localization}\\
\textbf{}& \textbf{(in degrees)} & \textbf{ ($z$) } & \textbf{($\beta$)} & \textbf{(egrs cm$^{-2}$ sec$^{-1}$)} & \textbf{error radius (in degreees)} \\
\hline 080825C & 60 & --   & -2.34$\pm$0.09 & (9.4$\pm$4)$\times10^{-6}$     & 0.95\\
\hline 080916C & 48 & 4.35 & -2.08$\pm$0.06 & (7.0$\pm$1)$\times10^{-5}$     & 0.09\\
\hline 090323  & -- & 3.57 & --           & (3.6$\pm$0.8)$\times10^{-5}$   & 0.09\\
\hline 090328  & -- & 0.74 & -2.86$\pm$0.10 & (3.2$\pm$2)$\times10{-5}$     & 0.21\\
\hline 090510  & -- & 0.90 & -2.60$\pm$0.30 & (3.7$\pm$0.7)$\times10^{-5}$   & 0.10\\
\hline 090626  & 15 & --   & -2.98$\pm$0.02 & (9.6$\pm$6)$\times10^{-6}$     & 0.21\\
\hline 090902B & 52 & 1.82 & -3.85$\pm$0.25 & (5.9$\pm$0.6)$\times10^{-4}$   & $<$0.14\\
\hline 091003  & 13 & --   & -2.64$\pm$0.24 & (1.3$\pm$0.8)$\times10^{-5}$   & $<$0.25\\
\hline
\end{tabular}
\end{center}
\end{table*}

\section{Conclusions:}

The estimation of the LAT localization errors as a function of the brightness of GRBs are presented. 
The analysis indicates the importance of ground-based robotic optical facilities for the 
multi-wavelength observations of these rare events. ROTSE-III observations of GRB 090902B indicate 
that the optical afterglow was one of the brightest at early epochs. Other LAT GRBs with measured 
redshifts, observed so far starting $\sim$16 hours after the burst, share the typical optical 
brightness seen in case of Pre-Swift/Swift GRBs. More GRBs detected at the LAT energies and their 
follow-up observations will shed light on the early temporal properties of the afterglows and 
their possible correlation with the observed GeV emission.


\begin{acknowledgments}

This research has made use of the data obtained through the High Energy Astrophysics Science Archive 
Research Center Online service, provided by the NASA/Goddard Space Flight Center. The ROTSE project 
is supported by the NASA grant NNX08AV63G and the NSF grant PHY-0801007.

\bigskip 

\end{acknowledgments}


\begin{thebibliography}{99} 
\bibitem{Dingus1995} Dingus B. L., 1995, Ap\&SS, 231, 187
\bibitem{Band2009} Band D., et al. 2009, ApJ, 701, 1673
\bibitem{Akerlof1999} Akerlof C. W. et al., 1999, Nature, 398, 400
\bibitem{Vestrand2006} Vestrand W. T. et al., 2006, Nature, 442, 172
\bibitem{Akerlof2006} Akerlof C. W. \& Yuan F, 2006, [{\tt arXiv:0702295}]
\bibitem{Ghiselini2009} Ghiselini et al, 2009, Accepted to MNRAS, [{\tt arXiv:09102459}]
\bibitem{Rossi2009} Rossi E. et al., 2009, \emph{The poster presented in this meeting}
\bibitem{Bissaldi2009} Bissaldi E., \& Connaughton V., 2009, GCN Circ., 9866
\bibitem{Kennea2009} Kennea, J., \& Stratta, G., 2009, GCN Circ., 9868
\bibitem{Quimby2006} Quimby, R. M., et al., 2006, ApJ, 640, 402 
\bibitem{Pandey2009} Pandey, S. B., Zheng, W., et al., 2009, GCN Circ., 9878
\bibitem{Abdo2009} Abdo A. A. et al., 2009, ApJ, 706, L138 
\bibitem{Akerlof2007} Akerlof C. W. \& Swan H. F., 2007, ApJ, 671, 1868
\bibitem{Guidorazi2009} Guidorazi C. et al., 2009, GCN Circ. 9875
\bibitem{Kann2007} Kann, A. et al., 2007, Submitted to ApJ, [{\tt arxiv:0712218}]
\bibitem{Haislip2009} Haislip J. et al., 2009, GCN Circ. 10003
\end{thebibliography}
\end{document}